\documentclass[12pt]{article}
\title{Special functions and the Mellin transforms of Laguerre and Hermite functions}
\author{Mark W. Coffey\\
Department of Physics\\
Colorado School of Mines\\
Golden, CO  80401\\
(Received $\mbox{~~~~~~~~~~~~~~~~~~~~~~~~~~~~~~~2006}$)}
\date{April 23, 2006}
\begin{document}
\maketitle
%\vspace{.25cm}
\baselineskip=25 pt
\begin{abstract}

We present explicit expressions for the Mellin transforms of Laguerre and Hermite
functions in terms of a variety of special functions.  We show that many of
the properties of the resulting functions, including functional equations and
reciprocity laws, are direct consequences of transformation formulae of
hypergeometric functions.  Interest in these results is reinforced by the fact
that polynomial or other factors of the Mellin transforms have zeros only on the
critical line Re $s = 1/2$.  We additionally present a simple-zero 
Proposition for the Mellin transform of the wavefunction of the $D$-dimensional
hydrogenic atom.  These results are of interest to several areas including quantum mechanics and analytic number theory.

\end{abstract}

%\vspace{.25cm}
\vfill
\baselineskip=15pt
\centerline{\bf Key words and phrases}
\medskip

\noindent
Mellin transformation, Hermite polynomial, associated Laguerre polynomial, 
hypergeometric series, transformation formulas, generating function, reciprocity law, 
recursion relation, functional equation 
%Whittaker function, Bessel function, parabolic cylinder function

\pagebreak
\medskip
\vspace{.25cm}
\centerline{\bf PACS classification numbers}
02.30.Gp, 02.30.Uu, 02.30.-f

\bigskip
%\vfill
\centerline{\bf AMS classification numbers}
33C05, 33C15, 42C05, 44A15, 44A20

\bigskip
\centerline{\bf Author contact information}
fax 303-273-3919, e-mail mcoffey@mines.edu

\baselineskip=25pt
\pagebreak
\medskip
\centerline{\bf Introduction}
\medskip

The work of Bump et al. \cite{bumpng,bumpchoi,kurlberg} on a property of the
zeros of the Mellin transforms of Hermite and associated Laguerre functions
has created significant interest.  Since the zeros of these functions lie on
the critical line Re $s=1/2$, suggestions of connections with the Riemann
hypothesis have been made.  The Mellin transforms of orthogonal polynomials have
additional properties and some of these have been pointed out 
\cite{bumpng,bumpchoi}.

In this paper a complementary point of view is developed through the theory of
special functions.  We obtain explicit results with the aid of, and in terms of,
a variety of special functions.  In particular, when the Mellin transforms
are written in terms of the Gauss hypergeometric function, well known 
transformation formulae yield functional equations, reciprocity laws, and other
properties.  

Although many of our final results recover those found in Refs. \cite{bumpng}
and \cite{bumpchoi}, our approach may be useful in finding clues to other
algebraic connections, since, for instance, the harmonic oscillator Hamiltonian
may be given a group theoretic interpretation via the Weil representation of
SL$_2$(R) \cite{kurlberg}.  %[[further elaboration ??...
%mention appearance of Whittaker function .....]]
Our work is related to the development of extended theta function representations
of the Riemann zeta function $\zeta$.  We have very recently shown how to 
correspondingly generalize the important Riemann-Siegel integral formula
\cite{coffeypla}.  Bump and Ng \cite{bumpng}
and Keating \cite{keating} have shown how to generalize Riemann's second proof of
the functional equation of $\zeta(s)$ by using Mellin and Fourier transforms
of Hermite polynomials.  %more later on Keating's work--to look at
These types of results extend to $L$-functions and automorphic forms 
\cite{bumpchoi,kurlberg}.  As we encounter the classical Whittaker function,
there are analogs in a more general automorphic context.  Much of this
appears to have an interpretation in terms of group representations. %reword
%future effort:  to work out Keating to Bump, Ng et al. correspondences ...

We recall that the eigensolution of the fundamental quantum mechanical 
problems of the harmonic oscillator and hydrogenic atoms contain Hermite or
associated Laguerre polynomials, depending upon the coordinate system used and
the spatial dimension (e.g., \cite{nieto,coffeyjpa}).  Two quantum numbers appear
for indexing the energy levels and the angular momentum.  These wavefunctions
have a wide variety of applicability, including to image processing and the
combinatorics of zero-dimensional quantum field theory \cite{coffeyjpa}.  
Very recently additional analytic properties of these "quantum shapelets" have
been expounded \cite{coffeyjpa}.  The one-dimensional Coulomb problem has recently
reappeared as a model in quantum computing with electrons on liquid helium
films \cite{platzman,nieto2}.  %of UCSD--refs. to supply here .....
In addition, given the self-reciprocal Fourier transform property of Hermite
polynomials, there are several applications in Fourier optics 
\cite{coffey1994,horikis}.
Hermite and Laguerre polynomials are also important in random matrix theory.

The simple harmonic oscillator and Coulomb problems may be transformed to
one another.  For example, the 4-dimensional (8-dimensional) harmonic oscillator
may be transformed to a 3-dimensional (5-dimensional) Coulomb problem 
\cite{ks,lambert}.  Many more mappings are realizable, especially for the
corresponding radial problems \cite{ks}.  Therefore, the eigensolutions are
closely related for these two problems with central potentials.
%skip I guess mention exp gen func for H's and L's has appl. to the combinatorics
%of 0-dimensional quantum field theory ...

We specifically call attention to Ref. \cite{atak} wherein Mellin transforms
were calculated for hypergeometric orthogonal polynomials and relations
within the Askey scheme \cite{koe} were discussed.  However, that reference did
not consider the location of the complex zeros of the transforms or connections
to the integral representations of zeta functions.  Moreover, given a family
of orthogonal polynomials $p_n(x,\{\alpha_i\})$ where $\alpha_i$ are real
parameters, orthogonal with respect to a weight function $w(x,\{\alpha_i\})$,
we are most interested in the Mellin transform of the associated functions
$\sqrt{w(x,\{\alpha_i\})}p_n(x,\{\alpha_i\})$.  This manner of including the
weight function is very useful in identifying the orthogonality and other
properties of the Mellin transforms, and was often not the case in Ref.
\cite{atak}.

%next give some remarks on organization of paper--to be refined later ...  
%pm:  start putting in the references
We first consider the Mellin transformation of Laguerre functions, outlining
multiple proofs of a result in terms of the hypergeometric function $_2F_1$.
We then briefly present similar results for the Mellin transforms of Hermite
functions.  Corollaries of our Propositions include the functional equations and reciprocity laws at negative integers of polynomial or other factors of the
transforms.  Other results include generating functions and recursion relations of
the Mellin  transforms.  We also strengthen a zero result of Ref. \cite{bumpchoi}.
We give a simple-zero result for the Mellin transform of the wavefunction for
$D$-dimensional hydrogenic atoms and then supply some concluding remarks.

\medskip
\centerline{\bf Mellin transform of associated Laguerre functions}
\medskip

We put for the Laguerre functions ${\cal L}_n^\alpha(x)=x^{\alpha/2}e^{-x/2}
L_n^\alpha(x)$ for $\alpha>-1$,
where $L_n^\alpha$ is the associated Laguerre polynomial \cite{andrews,grad,lebedev}.
We also put for the Mellin transform
$$M_n^\alpha(s) \equiv \int_0^\infty {\cal L}_n^\alpha(x) x^{s-1} dx=2^{s+\alpha/2}
\Gamma(s+\alpha/2)P_n^\alpha(s).  \eqno(1)$$
This transform exists for Re $(s+\alpha/2)>0$.
{\newline \bf Proposition 1}  We have
$$P_n^\alpha(s) = {{(1+\alpha)_n} \over {n!}} ~_2F_1(-n,s+\alpha/2;\alpha+1;2),
\eqno(2)$$
where $(a)_n=\Gamma(a+n)/\Gamma(a)$ is the Pochhammer symbol and $\Gamma$ is the
Gamma function.

\noindent
{\bf Corollary 1}  The polynomials $P_n^\alpha(s)$ satisfy the functional equation
$$P_n^\alpha(s)=(-1)^n P_n^\alpha(1-s).  \eqno(3)$$
{\bf Corollary 2}  The polynomials $P_n^\alpha(s)$ satisfy the reciprocity law
$${{(1+\alpha)_m} \over {m!}}P_n^\alpha(-m-\alpha/2)={{(1+\alpha)_n} \over {n!}}P_m^\alpha(-n-\alpha/2).  \eqno(4)$$
{\bf Corollary 3}  The polynomials $P_n^\alpha(s)$ have derivative
$${d \over {ds}}P_n^\alpha(s)={{(1+\alpha)_n} \over {n!}}\sum_{k=0}^n {{(-n)_k} 
\over {(1+\alpha)_k}}(s+\alpha)_k {2^k \over {k!}} \sum_{j=0}^{k-1} {1 \over 
{s+\alpha/2+j}}.  \eqno(5)$$
{\bf Corollary 4}  The polynomials $P_n^\alpha(s)$ may be written as
$$P_n^\alpha(s)={{(1+\alpha)_n} \over {n!}}\sum_{k=0}^n {{(-n)_k} 
\over {(1+\alpha)_k}} {2^k \over {k!}} \sum_{j=0}^k (-1)^{k+j} s(k,j)(s+\alpha/2)^j,
\eqno(6)$$
where $s(k,j)$ are Stirling numbers of the first kind.

Equation (3) has the form $\xi(s)=a \xi(1-s)$ where $|a|=1$.  Such a functional
equation is standard in the theory of completed zeta functions.  Corollary 1
follows from Proposition 1 upon the use of the transformation formula \cite{grad}
$$_2F_1(\alpha,\beta;\gamma;z)=(1-z)^{-\alpha} ~_2F_1\left (\alpha,\gamma-\beta;\gamma;
{z \over {z-1}}\right ).  \eqno(7)$$
Corollary 2, that has a combinatorial interpretation from counting lattice points
\cite{bumpchoi}, follows from Proposition 1 from the obvious symmetry
$_2F_1(-n,-m;\alpha+1;2)=$ $_2F_1(-m,-n;\alpha+1;2)$.
Corollary 3 follows from Proposition 1 by using the series definition of $_2F_1$
and the derivative of the Pochhammer symbol in terms of the digamma function 
$\psi \equiv \Gamma'/\Gamma$.  We have
$${d \over {ds}}P_n^\alpha(s)={{(1+\alpha)_n} \over {n!}}\sum_{k=0}^n {{(-n)_k} 
\over {(1+\alpha)_k}}(s+\alpha)_k {2^k \over {k!}}[\psi(s+\alpha/2+k)-\psi(s+\alpha/2)].
\eqno(8)$$
Applying the functional equation of the digamma function \cite{grad} then gives
Eq. (5).  The same method may be used to obtain higher order derivatives of 
$P_n^\alpha(s)$.  The series representation (6) for $P_n^\alpha(s)$ follows from
Proposition 1 by the series definition of $_2F_1$ and the expression for a
Pochhammer symbol in terms of $s(k,j)$ \cite{nbs}. 

In fact the polynomials of Eq. (2) are closely connected with the symmetric
Meixner-Pollaczek polynomials $P_n^{(\lambda)}(x,\pi/2)$ \cite{atak,koe}.
The latter polynomials are a special case of $P_n^{(\lambda)}(x,\phi)$ where
$\lambda >0$ and $0 < \phi < \pi$.  For the polynomials of Eq. (2) we identify
$P_n^\alpha(s)=(-i)^n P_n^{((1+\alpha)/2)}(i/2-is,\pi/2)$.

%somewhere give the measure wrt the P's are orthogonal ----------
For Proposition 1 we describe six separate proofs.  These are in addition to the
equivalent series result of Ref. \cite{bumpchoi} presumably based upon the power
series of the associated Laguerre polynomials \cite{grad,lebedev} and termwise 
integration.  We believe that the alternative proofs may contain illuminating
intermediate results that may have other applications.  Additional proofs are
possible using, for example, other integral representations of $L_n^\alpha$.
As this work was being finished, we found that Eqs. (1) and (2) are effectively
given in Ref. \cite{magnus} (p. 245).  Our alternative proofs illustrate a variety
of analytic techniques and exhibit connections with other special functions of
mathematical physics, in line with the theme of this paper.

{\em First proof}.  We start with the definition of $M_n^\alpha$ and make a 
change of variable:
$$M_n^\alpha(s)=2^{\alpha/2+s}\int_0^\infty e^{-y} y^{\alpha/2+s-1}L_n^\alpha(2y)
dy.   \eqno(9)$$
We then apply the property \cite{erdelyi}
$$L_m^\beta(\tau x)=\sum_{n=0}^m {{\beta+m} \choose {m-n}}\tau^n(1-\tau)^{m-n}
L_n^\beta(x) \eqno(10)$$
at $\tau=2$, giving
$$M_n^\alpha(s)=2^{\alpha/2+s}(-1)^n\sum_{j=0}^n (-1)^j 2^j {{\alpha+n} \choose
{n-j}} \int_0^\infty e^{-y}y^{\alpha/2+s-1} L_j^\alpha(y)dy.  \eqno(11)$$
The integral on the right side of this equation exists for Re $(s+\alpha/2)>0$
and is given in terms of \cite{grad} (p. 844) $_2F_1(-j,\alpha/2+s;\alpha+1;1)$.
In turn, this function value may be reduced to the ratio of Gamma functions
\cite{grad} (p. 1042) by Chu-Vandermonde summation.  Making these substitutions, transforming the Gamma factors, and writing them in terms of Pochhammer symbols,
gives Eq. (1) with $P_n^\alpha(s)$ as stated in Eq. (2).

{\em Second proof}.  Here we start with an integral representation of the associated
Laguerre polynomials in terms of Bessel functions of the first kind $J_\alpha$ and
develop an intermediate integral representation of $M_n(s)$ in terms of a Whittaker
function $M_{\mu,\nu}$.  We have \cite{andrews} (p. 286) 
$${\cal L}_n^\alpha(x) = {e^{x/2} \over {n!}}\int_0^\infty t^{n+\alpha/2} J_\alpha
(2\sqrt{xt})e^{-t} dt, \eqno(12)$$
so that the interchange of integrations gives
$$M_n^\alpha(s)={1 \over {n!}}\int_0^\infty t^{n+\alpha/2} e^{-t} dt \int_0^\infty
x^{s-1}e^{x/2} J_\alpha(2\sqrt{xt})dx.  \eqno(13)$$
Since \cite{grad} (p. 720)
$$\int_0^\infty x^{s-1}e^{x/2}J_\alpha(2\sqrt{xt})dx={{\Gamma(s+\alpha/2)} \over
{\sqrt{t}\Gamma(\alpha+1)}}e^t \left(-{1 \over 2}\right)^{1/2-s}M_{s-1/2,\alpha/2}
(-2t), \eqno(14)$$
with a change of variable we have
$$M_n^\alpha(s)={{\Gamma(s+\alpha/2)} \over {n! \Gamma(\alpha+1)}}\int_0^\infty
y^{n+\alpha/2-1/2}M_{s-1/2,\alpha/2}(y)dy  \eqno(15)$$
$$={{\Gamma(s+\alpha/2)} \over {n! \Gamma(\alpha+1)}}(-1)^{n+\alpha/2-s+1}
2^{\alpha/2+s}\Gamma(n+\alpha+1) _2F_1(n+\alpha+1,1-s+\alpha/2;\alpha+1;2).
\eqno(16)$$
In the last step we used Ref. \cite{grad} (p. 859).  Examining the asymptotic 
form of $M_{\mu,\nu}(z)$ for large $|z|$ shows that formula 7.621.1 there is valid
for Re $s>-1/2$, not just for Re $s >1/2$.  The reduction of Eq. (16) to Eq. (1)
with (2) is accomplished by applying the transformation formula \cite{grad}
$$_2F_1(\alpha,\beta;\gamma;z) = (1-z)^{\gamma-\alpha-\beta}  ~_2F_1(\gamma-\alpha,\gamma-\beta;\gamma;z). \eqno(17)$$ 

{\em Third proof}.  From the power series of the Laguerre polynomials we have
$$M_n^\alpha(s)=(1+\alpha)_n\sum_{k=0}^n {{(-1)^k} \over {k!(n-k)!(1+\alpha)_k}}
\int_0^\infty x^{\alpha/2+s-1+k} e^{-x/2} dx.  \eqno(18)$$
We evaluate the integral in terms of the Gamma function.  We then manipulate the
series, using $1/(n-k)!=(-n)_k (-1)^k/n!$, and Eq. (2) follows.

{\em Fourth proof}.  We substitute into the definition of $M_n^\alpha(s)$ the
expansion \cite{lebedev} (p. 89) 
$$x^{s-\alpha/2-1}=\Gamma(s+\alpha/2)\Gamma(s-\alpha/2)\sum_{j=0}^\infty 
{{(-1)^j L_j^\alpha(x)} \over {\Gamma(j+\alpha+1)\Gamma(s-\alpha/2-j)}},
\eqno(19)$$
make a change of variable, and twice apply property (10).  We then use the
orthogonality of the associated Laguerre polynomials with respect to the
measure $x^\alpha e^{-x} dx$.

{\em Fifth proof}.  We make use of the connection with the confluent hypergeometric
function $_1F_1$ (e.g., \cite{lebedev}, p. 273),
$$L_n^\alpha(x)={{(1+\alpha)_n} \over {n!}} ~_1F_1(-n;\alpha;x).   \eqno(20)$$
We then use a tabulated integral \cite{grad} (p. 860), giving Eq. (2).  Thus, this 
method is among the most expedient in obtaining the desired Mellin transform.

{\em Sixth proof}.  By using the exponential generating function of $L_n^\alpha$
\cite{andrews} (p. 288), it is readily verified that \cite{bumpchoi}
$$\sum_{n=0}^\infty M_n^\alpha(s)t^n=2^{s+\alpha/2}\Gamma(s+\alpha/2)(1-t)^{s-\alpha/2-1}
(1+t)^{-s-\alpha/2}, ~~~~~|t| < 1.  \eqno(21)$$
By then using the generalized binomial expansion
$$(1-t)^{-a}=\sum_{n=0}^\infty {{(a)_n} \over {n!}} t^n = ~_2F_1(a,b;b;t), \eqno(22)$$
we have
$$\sum_{n=0}^\infty P_n^\alpha(s)t^n = ~_1F_0(\alpha/2-s+1;.;t)~_1F_0(\alpha/2+s;.;-t)
\eqno(23a)$$
$$=\sum_{n=0}^\infty \sum_{\ell=0}^n {{(\alpha/2-s+1)_\ell (\alpha/2+s)_{n-\ell}} \over
{\ell ! (n-\ell)!}} (-1)^{n-\ell} t^n.  \eqno(23b)$$
By comparing powers of $t$ on both sides of Eq. (23b) we have
$$P_n^\alpha(s)=\sum_{\ell=0}^n {{(\alpha/2-s+1)_\ell (\alpha/2+s)_{n-\ell}} \over
{\ell ! (n-\ell)!}} (-1)^{n-\ell} ~~~~~~~~~~~~~~~~~~~~~~~~~~~~~~~~~~~~~~~~~~~~\eqno(24a)$$
$$={{(-1)^n \Gamma(\alpha/2+n+s)} \over {n!\Gamma(s+\alpha/2)}} ~_2F_1(-n,1+\alpha/2-s;
1-\alpha/2-n-s;-1).  \eqno(24b)$$
We then apply the linear transformation \cite{magnus} (p. 48)
$$_2F_1(a,b,c;z)= {{\Gamma(c)\Gamma(c-a-b)} \over {\Gamma(c-a)\Gamma(c-b)}} z^{-a} ~_2F_1
\left(a,a-c+1;a+b-c+1;1-{1 \over z} \right)$$
$$+ {{\Gamma(c)\Gamma(a+b-c)} \over {\Gamma(a)\Gamma(b)}} (1-z)^{c-a-b} ~_2F_1\left(c-a,1-a;c-a-b+1;1-{1 \over z} \right).  \eqno(25)$$
The appearance of the prefactor $1/\Gamma(-n)$ before the second resulting $_2F_1$
and the existence of simple poles of the Gamma function at nonpositive integers
annuls the second term.  We then simplify the first term using
$\Gamma(-\alpha)/\Gamma(-\alpha-n)=(1+\alpha)_n$ and Eq. (2) follows.    

We remark that Eqs. (24) may be obtained as a highly degenerate case of a formula
$_2F_1(a,b;c;gz) ~_2F_1(\alpha,\beta;\gamma;hz)=\sum_{n=}^\infty c_k z^k$, where
the coefficients $c_k$ may be expressed in terms of a generalized hypergeometric
function $_4F_3$.  In our case, $b=c$, $\beta=\gamma$, and $g=-h=1$.

%On to RRs for $M_n^\alpha$ and $P_n^\alpha$--see pp. 6 and 7 of my 1/14/06 notes
We next present example recursion relations satisfied by $M_n^\alpha(s)$ and 
$P_n^\alpha(s)$.  We have
{\newline \bf Proposition 2}.  
$${1 \over 2}[M_n^\alpha(s)+M_{n+1}^\alpha(s)]=(\alpha/2+s-1)[M_{n+1}^\alpha(s-1)
-M_n^\alpha(s-1)], \eqno(26a)$$
$$\left[1+{1 \over n}\left({\alpha \over 2}+s\right)\right]M_n^\alpha(s)={1 \over {2n}}
M_n^\alpha(s+1)+\left(1+{\alpha \over n}\right)M_{n-1}^\alpha(s), \eqno(26b)$$
$$(s+\alpha/2)[P_n^\alpha(s)+P_{n+1}^\alpha(s)]=(\alpha/2+s-1)[P_{n+1}^\alpha(s-1)
-P_n^\alpha(s-1)], \eqno(27a)$$
and
$$\left[1+{1 \over n}\left({\alpha \over 2}+s\right)\right]P_n^\alpha(s)={1 \over n}
(s+\alpha/2+1)P_n^\alpha(s+1)+\left(1+{\alpha \over n}\right)P_{n-1}^\alpha(s).
\eqno(27b)$$

{\em Proof}. Equations (27) follow from Eqs. (26) upon using the definition of $P_n^\alpha(s)$ in Eq. 1.  Equation (26a) follows from the use of the relation
\cite{grad} (p. 1037)
$$L_n^\alpha(x)={d \over {dx}}[L_n^\alpha(x) - L_{n+1}^\alpha(x)]  \eqno(28)$$
in Eq. (1) and integration by parts.  Similarly, Eq. (26b) follows from the
relation \cite{grad} (p. 1037)
$$L_n^\alpha(x)={x \over n}{d \over {dx}}L_n^\alpha(x)+\left(1+{\alpha \over n}
\right)L_{n-1}^\alpha(x), \eqno(29)$$
and integration by parts.

Bump et al. \cite{bumpchoi} noted the orthogonality of the polynomials $P_n^\alpha
(1/2+it)$ with respect to the measure $2^{\alpha+1}|\Gamma(1/2+\alpha/2+it)|^2dt$.
By applying the Plancherel formula they demonstrated that all the zeros of $P_n^\alpha
(s)$ have real part $1/2$ (\cite{bumpchoi}, Theorem 4).  By applying classical
results \cite{szego} (Theorems 3.3.1-3.3.3), we may strengthen their conclusion to
{\newline \bf Proposition 3}.  (a) The zeros of $P_n^\alpha(s)$ are simple and lie on
the line Re $s=1/2$.  (b)  The zeros of $P_n^\alpha(1/2+it)$ and $P_{n+1}^\alpha
(1/2+it)$ separate each other.
%get page number/Theorem number of Szego to insert above ---TO DO---------------------

Proposition 3 also follows from Theorems 5.4.1 and 5.4.2 of Ref. \cite{andrews}.

%later, just expand by the gen. binomial thm one of/each of the three factors in
%the integrand on p. 4 of my 1/14/06 pm notes.

Due to the relations $L_n^{-1/2}(x)=(-1)^nH_{2n}(\sqrt{x})/2^{2n}n!$ and 
$L_n^{1/2}(x)=(-1)^nH_{2n+1}(\sqrt{x})$ $/\sqrt{x}2^{2n+1}n!$ \cite{lebedev} (p. 81), 
where $H_m$ are the Hermite polynomials, the theory of the Hermite polynomials 
may be deduced from that of the associated Laguerre polynomial $L_n^\alpha$ in
the cases that $\alpha = \pm 1/2$.  Nonetheless, we believe that some separate 
discussion for Hermite polynomials is in order.  We develop the hypergeometric
function representation of their Mellin transforms, relate these to other special
functions, and consider reciprocity relations for a factor of the Mellin transforms.

\pagebreak
%\medskip
\centerline{\bf Mellin transform of Hermite functions}
\medskip

Following the normalization of Ref. \cite{bumpng}, we put 
$$f_n(x)=(8\pi)^{-n/2}H_n(\sqrt{2\pi} x)e^{-\pi x^2}.  \eqno(30)$$
For the Mellin transform, we set
$$M_n(s)=2\int_0^\infty f_n(x) x^{s-1} dx=\pi^{-s/2}\Gamma(s/2)p_{n/2}(s),
~~~~~~~~\mbox{Re}~ s >0.  \eqno(31)$$
We first consider the case of even degree Hermite polynomials and have
{\newline \bf Proposition 4}.  
$$p_n(s)=(8\pi)^{-n} (-1)^n {{(2n)!} \over {n!}} ~_2F_1(-n,s/2;1/2;2).  \eqno(32)$$
\noindent
{\bf Corollary 5}  The polynomials $p_n(s)$ satisfy the functional equation
$$p_n(s)=(-1)^n p_n(1-s).  \eqno(33)$$
{\bf Corollary 6}  The polynomials $p_n(s)$ satisfy the reciprocity law
$$(8\pi)^{-m} (-1)^m {{(2m)!} \over {m!}}p_n(-2m)=(8\pi)^{-n} (-1)^n {{(2n)!} \over
{n!}} p_m(-2n).  \eqno(34)$$
{\bf Corollary 7}  The polynomials $p_n(s)$ may be expressed as
$$p_n(s)=(8\pi)^{-n} {{\Gamma(s/2-n)} \over {\Gamma(s/2)}} (2n)! C_{2n}^{s/2-n}
(\sqrt{2}), \eqno(35)$$
where $C_n^\lambda$ is the Gegenbauer (generalized Legendre) polynomial
\cite{grad,lebedev,magnus}.  

Corollary 5 again follows from the transformation (7).  Corollary 6, that may
not have been observed before, follows from the symmetry of $_2F_1$ when its
two numerator parameters are interchanged.  It would be of interest to know if
Eq. (34) has a combinatorial interpretation.  Corollary 7 follows from the relation
\cite{grad} (p. 1030)
$$_2F_1(-n,s/2;1/2;2)=(-1)^n {s \over 2}B(s/2-n,n+1)C_{2n}^{s/2-n}(\sqrt{2}),
\eqno(36)$$
where $B$ is the Beta function.  In this way, known recursion formulas for the
Gegenbauer polynomials may be used to obtain the corresponding ones for the
polynomials $p_n(s)$.  

A great many methods may be used to obtain Eq. (32).  We simply indicate two of
them here.

{\em Proof 1}.  If we use the relation \cite{grad} (p. 1033),
$H_{2n}(x)=(-1)^n (2n)! ~_1F_1(-n,1/2;x^2)/n!$, with a change of variable we may
use a tabulated integral \cite{grad} (p. 860).

{\em Proof 2}.  If we perform the Mellin transform directly, we have
$$M_{2n}(s)=2 (8\pi)^{-n}(2\pi)^{-s/2}\int_0^\infty y^{s-1}H_{2n}(y)e^{-y^2/2}dy$$
$$=2 (8\pi)^{-n}(2\pi)^{-s/2}{{(-1)^n} \over {\sqrt{\pi}}} 2^{2n-3/2-(s-1)/2}4^{s/2}
\Gamma(s/2)\Gamma(n+1/2) ~_2F_1(-n,s/2;1/2;2), \eqno(37)$$
where we used \cite{grad} (p. 838).  Equation (32) is obtained after simplification
and noting that $\Gamma(n+1/2)=\sqrt{\pi}(2n)!2^{-2n}/n!$.

The function $p_{n/2}(s)$ is no longer a polynomial when $n$ is an odd integer.
We now have
{\newline \bf Proposition 5}.  
$$p_{n+1/2}=2^{5/2}(8\pi)^{-n-1/2}(-1)^n (n+1/2) {{(2n)!} \over {n!}} {{\Gamma(s/2+1/2)}
\over {\Gamma(s/2)}} ~_2F_1(-n,s/2+1/2;3/2;2).  \eqno(38)$$
In this case, the reciprocity law relates $p_{n+1/2}(-2m-1)$ and $p_{m+1/2}(-2n-1)$.
The Mellin transform $M_{2n+1}(s)$ may be performed directly, with the aid of
\cite{grad} (p. 838).  While $p_{n+1/2}(s)$ is not a polynomial in $s$, the
truncating $_2F_1$ in Eq. (38) is.

Bump and Ng \cite{bumpng} (p. 197) gave a recursion relation for $p_k(s)$.  Many
more recursion formulae exist for $p_k(s)$ and $M_n(s)$ and we mention two such.
We have
{\newline \bf Proposition 6}
$$M_n(s)={1 \over {\sqrt{2\pi}}}M_{n-1}(s+1)-{{(n-1)} \over {4\pi}}M_{n-2}(s),
\eqno(39)$$
and
$$M_n(s)={1 \over {4(n+1)}}\left[M_{n+1}(s+1)-{{(s-1)} \over {2\pi}}M_{n+1}(s-1)
\right].  \eqno(40)$$
Equation (39) follows from the use of $H_n(u)=2uH_{n-1}(u)-2(n-1)H_{n-2}(u)$ in
Eq. (1) while Eq. (40) follows from the use of $H_n(u)=H_{n+1}'(u)/2(n+1)$ 
\cite{grad} (p. 1033).

Finally, we obtain the generating function of $M_n(s)$ and $p_{n/2}(s)$ and describe
how the functional equation (33) arises from that of the parabolic cylinder function
$D_\nu$.  By using the generating function of the Hermite polynomials \cite{lebedev}
(p. 60) we have
$$\sum_{n=0}^\infty {{M_n(s)} \over {n!}} t^n = 2(8\pi)^{-n/2} e^{-t^2} \int_0^\infty
e^{-\pi x^2+2\sqrt{2\pi}xt} x^{s-1} dx$$
$$=2(8\pi)^{-n/2}e^{(\pi-1)t^2} (2\pi)^{-s/2}\Gamma(s)D_{-s}(-2t),  ~~~~~~\mbox{Re}~
s > 0, \eqno(41)$$
where we used \cite{grad} (p. 337) to evaluate the integral.  Thereupon, from Eq. (31),
we have
$$\sum_{n=0}^\infty {{p_{n/2}(s)} \over {n!}} t^n = 2(8\pi)^{-n/2}e^{(\pi-1)t^2} {2^{s/2-1}\over {\sqrt{\pi}}}\Gamma\left({{s+1} \over 2}\right) D_{-s}(-2t),
\eqno(42)$$
where the duplication formula of the Gamma function has been used.
The important functional equation (33) is recovered from this equation by using that
of $D_{-s}$ \cite{magnus} (p. 325).  Another way to see this is to note the connection
with the confluent hypergeometric function \cite{magnus} (p. 324)
$$D_\nu(z)=2^{\nu/2}e^{-z^2/4}\left [{{\Gamma(1/2)} \over {\Gamma[(1-\nu)/2]}} ~_1F_1
(-\nu/2;1/2;z^2/2)+ {z \over \sqrt{2}}{{\Gamma(-1/2)} \over {\Gamma(-\nu/2)}} ~_1F_1
[(1-\nu)/2;3/2;z^2/2) \right ]  \eqno(43)$$
and that Kummer's first transformation $_1F_1(\alpha;\rho;z)=e^z ~_1F_1(\rho-\alpha;
\rho;-z)$ applies.

\medskip
\centerline{\bf Mellin transform of the solution for hydrogenic atoms in $D$-dimensions}
\medskip

We now return to some of the quantum mechanical considerations of the Introduction.
We let $\psi(x)$ be the wavefunction for the hydrogenic atom (Coulomb problem) in
$D$-dimensions, where $x=(x_1,\ldots,x_D)$ and $r=|x|$.  We then have
{\newline \bf Proposition 7}.  The Mellin transform
$$\int_{R^D} \psi(x) r^{s-D/2-1}dx  \eqno(44)$$
has zeros only on the critical line Re $s=1/2$ and these zeros are simple.

{\em Proof}.  For hydrogenic atoms of nuclear charge $Ze$, the scaled Hamiltonian
is given by $H=-\nabla^2 + V(r)$ with potential energy $V(r) = -Ze^2/r$, where
$e$ is the electronic charge.  The eigensolutions satisfy $H\psi_{n\ell}=E_n
\psi_{n\ell}$, where $n \geq 1$ is the principal quantum number, $\ell=0,1,\ldots,n-1$ the angular momentum quantum number, and $E_n \propto -1/\eta^2(n)$ the energy levels, with $\eta(n) \equiv n + (D-3)/2$.  
(The energies are degenerate, meaning that they are independent of $\ell$ here.)
%these are degenerate to lowest order
For the central potential, the wavefunction is separable, so that
$\psi(x)=Y R_{n\ell}(r)$, where the function $Y$ is independent of $r$, and 
explicitly we have \cite{nieto}
$$R_{n\ell}(r)=\kappa_{n\ell} r^\ell e^{-r/\eta(n)}L_{n-\ell-1}^{2\ell+D-2}\left[{{2r}
\over {\eta(n)}}\right ], \eqno(45)$$
where the normalizing constant $\kappa_{n\ell}$ is independent of $r$.
%and $\eta(n) \equiv n + (D-3)/2$.
We have
$$\int_{R^D} \psi(x) r^q dx = \int_{S^{D-1}} Y d\Omega \int_0^\infty R_{n\ell}(r)
r^q r^{D-1} dr, \eqno(46)$$
where $S^{n-1}$ is the sphere in $R^n$ and $d\Omega$ the measure on it.
We put Eq. (45) into the radial integration of this equation and change variable
to $u= 2r/\eta(n)$.  We apply Proposition 3 with $q=s-1-D/2$ and Proposition 7
follows.

{\em Remarks}.  (i) We have just considered the standard Coulomb problem in $D$
dimensions.  However, it should be noted that only in three dimensions is the
potential $V(r) \propto 1/r$ the same as the Greens function of the Poisson 
equation $-\nabla^2 \phi = 4 \pi \rho$, where $\phi$ is the electrostatic potential
and $\rho$ the charge density \cite{nieto}.  (ii) The relativistic pi-mesic atom
is described by the Klein-Gordon equation.  This problem may be transformed to the
hydrogenic atom in any dimension, and the wave equation solutions are functionally
identical \cite{nieto}.  In particular, the same radial wavefunction (45) appears.
Therefore, Proposition 7 could instead be stated in terms of the wavefunction of
a pi-mesic atom.  (iii) The 3-dimensional hydrogenic atom is also separable in
parabolic coordinates $(\xi,\eta,\phi)$ \cite{schiff,bluhm}.  In this case the
wavefunction
$u_{n_1n_2m}(\xi,\eta,\phi) \propto e^{-\xi/2}\xi^{|m|/2}L_{n_1+m}^{|m|}(\xi) 
e^{-\eta/2}\eta^{|m|/2}L_{n_2+m}^{|m|}(\eta) e^{im\phi}$, $m=0$, $\pm 1$, $\pm 2, 
\ldots$, contains Laguerre functions in both of the confocal paraboloid
coordinates.

%\medskip
\pagebreak
\centerline{\bf Summary and brief discussion}
\medskip

We have explicitly evaluated the Mellin transforms of Hermite and associated
Laguerre functions in terms of several other special functions.
In particular, our approach permits the arsenal of known results for 
hypergeometric functions to be applied.  For instance, the transformation formulae and 
symmetries of the function $_2F_1$ directly lead to the functional equation and
reciprocity law of the polynomial factor of the Mellin transform.   The Hermite and
associated Laguerre functions and their Mellin transforms are of much interest in
analytic number theory due to their connections with generalizations of Riemann's
second proof of the functional equation of his zeta function 
\cite{bumpng,bumpchoi,kurlberg}.  This interest is reinforced since the polynomial
factors of the Mellin transforms have simple zeros that occur only on the critical
line Re $s=1/2$.   

We have described links of the Mellin transforms to fundamental problems of 
quantum mechanics, including the isotropic harmonic oscillator and Coulomb problems.
Our Proposition 7 evidences such connections for the solution of the Schrodinger
equation with central potential.  Classically, only for the harmonic oscillator and
Kepler problem are the orbits always closed.  Transformation between these two
problems in quantum mechanics was noted by Schrodinger himself and there is 
continuing interest in this subject.  In addition to our work, known algebraic
relations (e.g., \cite{lambert}) may yield additional insight into the group
representation aspects behind the Mellin transforms of interest.
%cite character sum analogs of the hypergeometric function and Hermite/Laguerre polys?
We also mention that character sum analogs over finite fields are known for the
Hermite polynomials \cite{evans}.

The orthogonality of factors of Mellin transforms of solutions of the Schrodinger
equation may be viewed as follows.  Since the Schrodinger equation with suitable 
boundary conditions is self-adjoint, its eigenfunctions corresponding to distinct
eigenvalues (eigenenergies) are orthogonal.  Then the isometry of the Fourier
(Mellin) transform takes the orthogonality in real space to that in momentum space.

%we can bring to bear hypergeometric function theory systematically with our results ...
%go over Keating paper--there is other stuff yet to be brought in here?
%cover open question on the funcs. w/ beta instead of -n in the 2F1 form ...

Our Proposition 1 suggests several questions.  The most begging question may be
the following.  If we put
$f(s) = ~_2F_1(\beta,s+\alpha/2;\alpha+1;2)$, Eq. (7) shows its functional equation
to be $f(s) = (-1)^\beta f(1-s)$.  Therefore, we ask what conditions must be placed
upon the parameter $\beta$ in order for $f$ to have zeros only on the critical line?
The development of an orthogonality relation for such functions with respect to an
appropriate measure would probably address this.  %go back and review Bump and Choi paper
%how about for multivariable hypergeometric functions?
%and what of the pFq that reduce to a polynomial--see, e.g., p. 64, Section 2.10 of
%Magnus et al.
At least two other hypergeometric extensions may be possible.  We first consider 
all cases of the extended hypergeometric function $_pF_q$ that reduce to 
polynomials (e.g., \cite{magnus}, p. 64) and correspond to a self-adjoint boundary
value problem.  We then ask which of these have a functional equation relating argument $z$ to $1-z$ and possess zeros only along Re $z = 1/2$.
Another direction is to consider multivariate hypergeometric functions.  This 
includes the two-variable Kamp\'{e} de F\'{e}riet function and its generalizations.

Another avenue of generalization would be through the connection of the extended
hypergeometric function to the Meijer $G$-function (\cite{grad}, p. 1071).  
In particular, interesting candidate functions are offered by the $G$-function at
argument $1$ or $-1$, as then the functional relations of $G$ show there are
functional relations at parameters $a_r$, $b_s$ and $1-b_s$ and $1-a_r$.  

Within the Askey scheme \cite{koe}, the Wilson polynomials $W_n(x^2;a,b,c,d)$
with Re $(a,b,c,d) > 0$ are orthogonal on $[0,\infty)$ with respect to the measure
$(2\pi)^{-1} |\Gamma(a+ix)\Gamma(b+ix)\Gamma(c+ix)\Gamma(d+ix)/\Gamma(2ix)|^2 dx$.
Therefore it seems worthwhile to consider in the future the Mellin transform of
the function
$${\cal{W}}_n(x^2;a,b,c,d)={1 \over  {\sqrt{2\pi}}}\left|{{\Gamma(a+ix)\Gamma(b+ix)
\Gamma(c+ix)\Gamma(d+ix)} \over {\Gamma(2ix)}}\right| W_n(x^2;a,b,c,d).  \eqno(47)$$

\bigskip
\centerline{\bf Acknowledgement}
This work was partially supported by Air Force contract number FA8750-06-1-0001.

\pagebreak

\end{document}